\documentstyle{elsart}

\newcommand{\text}{\rm}

\begin{document}

\begin{frontmatter}

\title{ Chaos and isospin symmetry breaking in rotational nuclei }
\author[Strasbourg,Munchen]{J.A. Sheikh},
\author[Debrecan]{A.T. Kruppa},
\author[Strasbourg]{N. Rowley}
\address[Strasbourg]{Institut de Recherches Subatomiques (IReS),
Universite Louis Pasteur,
23 rue du Loess, F-67037 Strasbourg Cedex 2, France}
\address[Munchen]{Physik-Department, Technische Universit\"at M\"unchen,
D-85747 Garching, Germany }
\address[Debrecan]{Institute of Nuclear Research, H-4001 Debrecan,
P.O. Box 51, Hungary}

\maketitle

\begin{abstract}
For nuclei with $N\approx Z$, the isospin degree of freedom
is important and, for deformed systems, rotational bands of 
different isospin may be expected at low excitation energies. We 
have investigated, in a simple model space, the influence of the isospin-breaking
Coulomb interaction on the degree of chaoticity of these rotational
bands. The statistical measures used rely on an analysis of
level-spacing distributions, which are extremely difficult to
measure experimentally. We show, however, that the overlap intergrals
between states of similar frequency reflect well 
the degree of chaoticity. This quantity is closely related to the 
experimentally more accessible $\gamma$-decay ``spreading width''.
\end{abstract}
\end{frontmatter}

\section{Introduction}

Random matrix theory (RMT) \cite{gmw98} has played an important 
role in our
understanding of quantum many-body problems. In classical
mechanics the
underlying nature of a physical system is revealed \cite{rei92}
by the behaviour
of its trajectories in phase-space. For regular motion, these trajectories
are stable against small perturbations of their initial conditions, whereas 
in the case of chaotic motion the trajectories show
hyperbolic instability and tend to fill the whole of the 
available phase space. In quantum-mechanics, it is not possible
to define such trajectories since the position and momenta 
coordinates cannot be defined simultaneously.
Consequently it is not possible to determine the 
motion of the system in the same way. RMT~provides a valuable connection 
between the spectral properties of a quantum system and its corresponding
classical motion. It has been demonstrated that the spectral distributions
of few-body quantum systems which follow the predictions of RMT 
have chaotic motion
in the corresponding classical phase-space, whereas distributions
which follow a Poissonian behaviour display
regular classical motion. This is referred to as the
Bohigas-Giannoni-Schmit conjecture \cite{bgs84} and is now 
accepted as a generic feature of quantum many-body systems.

The underlying motion of a classical 
system depends on its symmetries or, in other words, on the 
integrals of motion. For a system with a number of integrals of motion
equal to the number of degrees of freedom, the classical motion is 
completely regular \cite{rei92}. In quantum mechanics, the quantum number is
the analogue of the integral of motion and it has been shown that symmetry
breaking leads to chaotic motion. This has been clearly demonstrated in
the interacting boson model (IBM) \cite{aw91}. This model has three
dynamical symmetry limits and the restriction 
to a particular dynamical symmetry
corresponds to regular motion. It is the interpolation among the three
relevant symmetry groups which gives rise to chaos.

The purpose of the present work is to investigate how
the spectral distribution is modified by the
inclusion of a two-body Coulomb force which leads of course to
isospin symmetry breaking. Such an analysis has already been 
carried out for spherical nuclei \cite{mbes88,gw90,ob92}. 
In the present work, however, we shall
study deformed nuclei and investigate the effects of
the isospin symmetry breaking as a function of rotational 
frequency, following the recent observation of rotational bands 
with good isospin for $N\approx Z$ \cite{rud96}. 
In the absence of a Coulomb force the states may be labelled by the
isospin $T$ and we shall see that each individual set of states 
of a given $T$, displays chaotic dynamics. The introduction of
the Coulomb force, however, breaks this symmetry and the entire 
set of states must now be considered together. The final dynamics  
depends on the strength of the Coulomb interaction relative to the
cranking and deformation energies. For relatively small deformations
and low rotational frequencies, chaotic dynamics are restored to
the full set of states. However, for large deformations and high
frequencies a more regular dynamics may persist.

We have performed the statistical analysis of eigen-values
using the measures of Brody \cite{bro81}, Berry-Robnik \cite{be84} 
and the spectral-rigidity of Dyson and Mehta \cite{meh91}.
A similar analysis has been carried out in ref.\cite{kpr95}
for a single type of nucleon. Of course such analysis demand 
data which are impossible to obtain. We have, therefore,
investigated the spreading of an eigen-state at one frequency 
over the many other eigen-states to which it can $\gamma$ decay at
a lower frequency. This quantity, which is closely related to
the $\gamma$-decay spreading width is indeed shown to reflect the 
degree of chaoticity evaluated by the above statistical analysis.
 
The cranking hamiltonian is 
briefly described in section II and the theoretical approaches used to
investigate the statistical distributions are given in section III.
The spectral analysis of the eigen-values is presented
in section IV, overlap integrals in section V 
and the results are summarized in section VI.

\section{Rotational Model}

We employ a deformed shell-model hamiltonian, 
consisting of a cranked deformed
one-body term, $h^\prime$ and a scalar-isoscalar two-body delta-function
residual nuclear interaction 
\cite{srn90,she90,fsr94}. The one-body term is the familiar
cranked Nilsson mean-field potential which takes account of the
long-range part of the nucleon-nucleon interaction.
Thus
\begin{equation} \label{H}
H^\prime = h^\prime -g \delta(\hat R_1 - \hat R_2) + V_c
\end{equation}
where,
\begin{equation}\label{h'}
h^\prime = -4 \kappa { \sqrt { 4 \pi \over 5 }} Y_{20}  - \omega J_x.
\end{equation}
The strength of the two-body interaction is 
$G=g\int R^4_{nl}r^2 dr$ and the deformation
energy $\kappa$ is related to the deformation parameter $\beta$ through
\begin{equation}
\kappa \simeq 0.16\hbar\omega_0(N+3/2)\beta,
\end{equation}
 where $\hbar\omega_0$ is the harmonic oscillator frequency of the
 deformed potential and $N$ the quantum number of the major shell. 
For the case of $f_{7/2}$ shell, $\kappa$=1.75 approximately
corresponds to $\beta=0.16$. The last term in Eq. (1), $V_c$ is the
Coulomb interaction among the valence protons. We have employed the
emipirically obtained \cite{swi98} two-body Coulomb matrix-elements in the
$f_{7/2}$ shell and are given by $V_c(J)= 0.578, 0.486, 0.374$ and
$0.330~$MeV for $J=0,2,4$ and $6$, respectively.
In order to solve the eigen-value problem exactly,
we are limited to a small configuration space. As in the previous work,
the model space in the present analysis consists of a single j-shell.
We have diagonalized the hamiltonian (\ref{H}) exactly for neutrons and protons
in the $f_{7/2}$ shell. As the strengths of nn-, pp- and np-parts 
are identical, the hamiltonian  in the absence of $V_c$ is
invariant with respect to rotations in isospace, i. e.
\begin{equation}
{\cal R}H^\prime {\cal R}^{-1}=H^\prime,
\end{equation}  
where ${\cal R}$ defines a rotation in isospace, generated by the 
isospin operators $T_x,~T_y$ and $T_z$.
Furthermore, the hamiltonian (1) is invariant with respect to a 
spatial rotation
about the x-axis by an angle of $\pi$. As a consequence, the signature $\alpha$
is a good quantum number \cite{bf79}, which implies that the shell model
solutions represent states with the angular-momentum
$I=\alpha +2n$ ($n$ integer).

\section{ Statistical methods }

In this section, we shall briefly outline the statistical methods used 
in the present work.
For more details on these methods the reader is referred to ref. \cite{kpr95} and the
references therein. 

The level spacing is defined by
\begin{equation}
S_i=X_{i+1}-X_i\ ;\ \ i=1,2,\ldots\ \ \ .
\end{equation}
where $X_i$ are the unfolded energy levels
\begin{equation}
X_i=\bar N(E_i)\ \ \ i=1,2,\ldots N.
\end{equation}
Roughly speaking $S_i$ is the original level spacing 
$(E_{i+1}-E_i)$
in units of the local average level separation.
The unfolding of the spectra can be carried out
in a different way using
\begin{equation}
S_i=(E_{i+1}-E_i)\bar\rho(E_i)\ \ i=1,2,\ldots N-1.
\end{equation}
This unfolding of the spectra ensures that the average 
spacing in the series
$X_i$ is unity. In this way the fluctuation properties 
of the spectra
of different systems can be compared. In our investigations 
both types
of unfolding procedures led qualitatively to the same outcome.
The results in the rest of the paper were produced using
the mapping $(7)$.

The first spectral statistics we use is the
nearest-neighbour distribution $P(S)$. The quantity $P(S)dS$ 
gives
the probability that the nearest-neighbour of an arbitrarily
selected level $S_i$ lies in the interval $(S_i+S,S_i+S+dS)$.
From the finite set of $\{S_i\}_{i=1}^N$ only a histogram can 
be constructed
as an NND. In order to have good statistics the bin size of
the histogram is chosen to ensure that there are at least seven
spacings in each bin. We considered the NND in the interval
$S\in(0,2)$.

Spectral statistics show Poissonian or GOE
forms for integrable and fully chaotic systems. For the case of mixed phase
phase the analytic forms of these statistics are not known.
In the following we shall employ three parametrizations of 
Berry-Robnik \cite{be84}, Brody \cite{bro81}
and the spectral-rigidity of Dyson and Mehta \cite{meh91}.
The Berry-Robnik parametrization of NND is given 
\begin{eqnarray}
P_{BR}(q,S)=&\bar q^2&\exp (-\bar qS)\,{\rm erfc}(\frac{1}{2}
\sqrt {\pi}
qS)\nonumber \\
&+&(2q\bar q+\frac{1}{2}\pi q^3S)\exp(-\bar qS-\frac{1}{4} \pi q^2S^2)
\end{eqnarray}
where $0\leq q \leq 1$ is the measure of the chaotic region 
of the phase
space and $\bar q=1-q$.

The Brody distribution is given by
\begin{equation}
P_{B}(b,S)=(1+b)AS^b\exp(-AS^{1+b}),
\end{equation}
where
\begin{equation}
A=\Gamma\left (\frac{2+b}{1+b}\right)^{1+b},
\end{equation}
and $\Gamma$ denotes the usual gamma function.
For
fully chaotic and ordered systems $q=b=1$ and $q=b=0$, 
respectively.

The spectral statistics $\bar\Delta_3(L)$ measures the long-range 
correlation of the unfolded levels
\begin{equation}
\Delta_3(X,L)=\min {{1}\over{L}}\int ^{X+L}_X (N_u(E)-(AE+B))^2dE,
\end{equation}
where $N_u$ is the cumulative level density of the unfolded
levels $X_i$. We average $\Delta_3(X,L)$ over intervals $(X,X+L)$ to
get $\bar\Delta_3(L)$, as outlined in \cite {bgs84}.
For ordered systems
$\bar\Delta_3(L)=L/15$ and for fully chaotic ones 
$\bar\Delta_3(L)\approx
\ln(L)/\pi^2-3/4$ for $L>>1$.
The explicit calculation of the spectral rigidity was done with
the method described in \cite{bo84b}.

The spectral rigidity for a mixed statistics \cite{se85a} is given by 
\begin{equation}
\Delta_3(q,L)=\Delta_3^P(qL)+\Delta_3^{GOE}((1-q)L).
\end{equation}
The $b$ and $q$ parameters of the NND distributions and the
spectral rigidity are determined
by least-squares fits to the numerical results.
The errors in the best-fit values of these parameters were 
estimated
by the method of maximum likelihood and the use of constant
chi-squared boundaries as a confidence limit \cite{pr86}.

\section{ Results and discussions }

The cranked shell model calculations have been carried out for (4-neutrons+
4-protons) in the $f_{7/2}$ shell, $(f_{7/2})^4_\nu(f_{7/2})^4_\pi$.
We have used the 
$f_{7/2}$ shell for simplicity in carrying out the exact deformed shell
model calculations. For this problem the dimensionality of the
matrix to be diagonalized for $\alpha=0$ (favoured configuration) is 2468.
In order to check the 
deformation dependence of the statistical distributions, we have performed 
calculations at two deformation values with $\kappa=1.75$ and $3.50$. Since
the cranking is a rather poor approximation at lower rotational frequencies,
most of the statistical analysis will be provided at higher frequencies.

The nearest neighbour distribution is shown in Fig. 1 for 
$\hbar \omega = 0.5$ MeV. The statistical analysis has been carried 
out for each isospin bin separately and the obtained NND clearly fits 
GOE for T=0,1 and 2 separately. For 
other isospin values (T=3 and 4), the number of states are very few and it is not 
possible to perform the statistical analysis. In fact for T=4, there is only one state.
In Fig. 2, the statistical analysis is performed with the same
parameters as in Fig. 1, but with all the 2468 states
without considering the isospin quantum number. 
Fig. 1(a) shows NND with no Coulomb interaction, isospin conserved. 
The obtained distribution is closer to the Poission than to Gaussian orthogonal
ensemble. 
This feature of obtaining a Poission like distribution from the superposition
of several GOE distributions has already been demonstrated in 
ref.\cite{gw90}. This can be explained by noting that since
there is no interaction between the states with good T, the near degeneracy
of two eigen-values belonging to two different isospin quantum numbers
is unavoidable and consequently gives rise to a more
Poission-like distribution which favours energy levels with zero spacings.

The NND in the presence of the Coulomb interaction is shown in Fig. 2(b) and
as is seen that the distribution becomes closer to GOE. The isospin 
mixing resulting from the Coulomb interaction is quite small and is 
shown in Table 1 for the lowest five-states and the 14th and the 15th states. 
We have obtained the
isospin mixing by expanding the wavefunction in terms of the states
with good T. The isospin mixing for the
lowest state at $\hbar \omega = 0.5$ MeV is about 0.02\%. The mixing 
for the lowest three T=0 states is from the T=1 state and components of the
other isospin states are negligible.
We would like to mention here that isospin mixing for some of the states,
for instance for the 14th and the 15th states
is quite large and is due 
to the near degeneracy of the two-states, the energies of the two-states being
-33.43 and -33.41, respectively. The isospin mixing in the present model analysis
appears to be quite small and is due to the fact that we consider 
Coulomb interaction
only among the valence protons. In principle, there should be also 
a contribution from the core protons.

In Fig. 2(c), the results are presented
by artificially increasing the Coulomb matrix elements by a factor of two and 
the distribution turns out to be more close to GOE with substantially
reduced chi-square. As one expects, for a strong symmetry
breaking the distribution should converge to a single GOE \cite{gw90}. Fig. 1
clearly demonstrates a strong dependence of the statistical distribution
on the residual Coulomb interaction.

As already mentioned, in the above statistical analysis, we have used all
the eigen-values. It is also of interest to perform the statistical 
analysis of different
excitation energy windows. In Fig. 3, the NND is plotted for three
bins each consisting of 800 eigen-states, the top panel shows NND
for the lowest 800 states. The middle panel from 801 to 1600 and the
lower panel from 1601 to 2400. As is evident from this figure, a strong
dependence of NND on excitation energy is obtained. This conforms with
earlier studies that the degree of chaoticity increases with the excitation
energy.

In Figs. 4, similar results as in Fig. 2 are presented but for 
$\hbar \omega =2.5$ MeV.
As compared to $\hbar \omega = 0.5$ MeV, the distributions now deviate more
from GOE and can be understood from the fact that with increasing frequency
the quasiparticles align along the axis of rotation and the
motion tends to be more regular. In order to quantify the deviations from
GOE statistics, we have analyzed the NND with Brody,
Berry-Robnik and spectral-rigidity parametrizations. The results are presented
in Fig. 5 with all the three measures. In Fig. 5(a), the parameters are 
presented with no Coulomb potential, including all the isospin states. The 
Brody parameter is small indicating that NND is regular for all 
the rotational frequencies studied. The Berry-Robnik and the spectral-rigidity
parameters on the other hand are between those expected for pure 
GOE and Poission distributions. In the presence of the Coulomb potential,
the results are given in Fig. 5(b) and as is apparent all the three measures 
give values close to 1 signifying that the underlying classical
motion is chaotic. We have also calculated these statistical measures by
considering the isospin bins separately and in Fig. 5(c), the results
are given for T=0. It is noted from Fig. 5(c) that all the three measures
give values very close to 1. Comparing the middle and the lower panels of
Fig. 5, it is observed that the degree of chaoticity is larger for T=0
level statistics than with Coulomb interaction. This unexpected result
can be understood in the following manner: The chaoticity for the T=0
states is determined by the residual interaction and in the present model
includes the neutron-proton interaction. The combined residual interaction is
quite strong and we obtain
chaotic behaviour. On the other hand, in the statistical analysis of 
Fig. 5(b), the 
degree of chaoticity is determined by the Coulomb interaction among the
good isospin states and in the present model this interaction is small
and consequently the distribution is less chaotic. This would also explain
the apparent contradiction with the results obtained in ref.\cite{mbes88}.
It was shown in ref.\cite{mbes88} that the distributions are similar 
whether one separates the experimental levels into isospin bins or not.
The realistic isospin mixing is stronger than calculated in the present
work and is the reason for the apparent discrepancy between the present work
and the results presented in ref.\cite{mbes88}.

The isospin mixing for $\hbar \omega =2.5$ MeV for the lowest few 
eigen-states are given Table 1. As compared to $\hbar \omega =0.5$ MeV, it is
observed that isospin mixing is lower. This can be explained by noting that
with increasing frequency the particle alignment occurs along the axis of
rotation and there is a transition from the paired $(J=0)$ configuration to
the aligned state $(J=2j-1)$ at the first bandcrossing. The Coulomb energy
is maximum for the paired state and is least for the aligned configuration
\cite{swi98} and this is the reason for the lowering of the isospin mixing with
increasing rotational frequency.

The spectral-rigidity statistical measure is provided in Fig. 6 for the three
cases studied earlier at $\hbar \omega = 0.5$ MeV. The results with
no Coulomb interaction and considering all the states, shown in the top
panel of Fig. 6 , are between Poission and GOE. The results with 
Coulomb interaction, shown in the middle panel, are close to GOE and agree
with the other statistical measures presented earlier. The results for T=0
levels shown in Fig. 6(c) exactly follow the expected GOE spectral-rigidity.

The results with larger deformation, $\kappa=3.5$ for 
Brody, Berry-Robnik and spectral-rigidity parametrizations are presented
in Fig. 7. It is noted that the degree of chaoticity is slightly less
as compared to $\kappa=1.75$ and can be understood from the fact that
with increasing deformation the collective effects dominate and the
results tend to become more regular.

\section{Overlap-integrals and the decay probabilities}

In a regular regime, it is expected that the wavefunction should evolve 
smoothly as a function of the cranking frequency. This is, indeed, the
case for the yrast wavefunction. The calculation of the overlap between
the cranking yrast wavefunctions at the neighbouring frequency points 
is almost equal to one, except in the bandcrossing region. However, in
the region of chaoticity, the overlaps are expected to have a wide spread.
This is illustrated in Fig. 8, where the overlaps are plotted between
$\hbar \omega =0.55$ MeV and $\hbar \omega =0.50$ MeV at three
different energy regimes. This overlap is proportional to the transitions
between the neighbouring spin states. In Fig. 8(a), the overlap is presented
for the 10th state which is close to the yrast line and it is clear that
it has a peak value for one state, for all other states the overlap is 
close to zero. The overlap for the 500th eigen-state is shown in Fig. 8(b)
and the results indicate that the overlap for this state has a broad
spectrum. This state lies in the regime of chaoticity and demonstrates that
the state cannot be traced as a function of the cranking frequency. 
In Fig. 8(c), the overlap-integral is plotted for the 1500th eigen-state and
the state appears to have a much more spread as compared to Fig. 8(b).
This can be easily explained by noting that the degree of chaoticity
increases with the excitation energy. 

It is also of interest to check
the overlap-integral for the case where the statistical analysis
shows regular motion at all excitation energies. In Fig. 9, the results
are presented for $\kappa=3.50$ and $\hbar \omega=2.5$ MeV which has a more
regular behaviour as is evident from Fig. 7(b). Fig. 9 shows that for all the three 
eigen-states considered at different excitation energies, the 
overlap-integral peaks for one particular bra eigen-state. From the
comparison of Fig. 8 and 9, it is clear that the degree of chaoticity and 
overlap-integal are closely related to each other. In Fig. 9, where
the statistical analysis shows chaotic features at high excitation
energy, the overlap-integral for a state in that regime shows a very 
high degree of fragmentation. In comparison, the overlap-integral
peaks for one particular state at the three chosen excitation energies.
For this chosen deformation
and cranking frequency parameters, the corresponding statistical
analysis indicate a more regular motion at the three considered excitation energies.

\section{Summary}

In the present work, the statistical distributions of the rotational bands
has been studied in the presence of the Coulomb potential. As already 
mentioned in the introduction, this is relevant to the recent
discovery of the isospin rotational bands near N=Z. The analysis
has been carried out using a simple cranked shell model approach. Although,
this model is not very realistic to make a direct comparison with the
experimental data, but it contains all the essential ingredients of a more 
realistic deformed rotational model. The advantage in the 
simple model is that it
can be solved exactly and the effects of the Coulomb interaction can
be evaluated precisely. In a realistic approach, Hartree-Fock or
Hartree-Fock-Bogoliubov many-body techniques are employed. It is known
that these approximations also contribute to the isospin mixing and it
is not possible to obtain a true measure of the isospin mixing.

We have shown that NND is very sensitive to the Coulomb interaction and 
can provide a measure of the isospin mixing in nuclei. The NND with the
inclusion of all the good isospin states (in the absence of the Coulomb 
potential) is closer to the Poissonian distribution. However, the
investigation of the NND for each isospin bin separetely, shows that 
the distribution is GOE. The degree of chaoticity for each isospin
bin is determined by the two-body residual interaction which in the
present work is very strong and we obtain NND which fits the distribution
of GOE. The presence of the Coulomb interaction among the good isospin states 
immediately gives rise to GOE-like distribution and it has been shown
that the degree of chaoticity is quite sensitive to the strength of the
Coulomb interaction.

It has been demonstrated that the degree of chaoticity depends on 
the deformation.
The chaotic features appear more prevalent in normal deformed bands 
as compared
to the superdeformed shapes.

Furthermore, it has been shown in section V that the overlap-integal
between the cranking wavefunctions at neighbouring frequencies
and the degree of chaoticity are 
related to one another. This overlap-integral is proportional to the
transition probabilities between the adjacent states. It has been demonstrated 
that in a regime where the statistical analysis shows chaotic motion, the corresponding 
overlap-integral is highly fragmented and for the regular motion the
overlap-integral peaks for one particular state. 

In a more realistic problem, this overlap would have to be taken between states of 
neighbouring spin. 
This would be I and I-2 for stretched E2 transitions. We are presently calculating
these transitions using the projected shell model \cite{hs} approach with a 
realistic interaction and model space. 



\newpage
\begin{table}
\caption{ The expansion coefficients of the lowest few
eigen-states with Coulomb-
interaction in terms of the 
states with good isospin. This gives a measure of the isospin mixing.
The calculations are presented with deformation parameter $\kappa=1.75$.
The results are also provided for the two lowest states which show very
large isospin mixing and is due to the near degeneracy of the corresponding
eigen-values.}
\begin{tabular}{ccccccc}
State &T=0     & T=1     &     T=2 & T=3 & T=4\\ 
\hline
&&&$\hbar \omega=0.5$ MeV &&\\
\hline
1& 0.99981 & 0.00015 & 0.00004 & 0.0 & 0.0 \\
2& 0.99797 & 0.00199 & 0.00003 & 0.0 & 0.0 \\
3& 0.99412 & 0.00587 & 0.00001 & 0.0 & 0.0 \\
4& 0.00573 & 0.99422 & 0.00005 & 0.0 & 0.0 \\
5& 0.99120 & 0.00836 & 0.00002 & 0.0 & 0.0 \\
14&0.12958 & 0.87026 & 0.00015 & 0.0 & 0.0 \\
15&0.87250 & 0.12740 & 0.00010 & 0.0 & 0.0 \\
\hline
&&&$\hbar \omega=2.5$ MeV &&\\
\hline
1& 0.99988 & 0.00011 & 0.00000 & 0.0 & 0.0 \\
2& 0.99951 & 0.00048 & 0.00000 & 0.0 & 0.0 \\
3& 0.99885 & 0.00115 & 0.00000 & 0.0 & 0.0 \\
4& 0.00223 & 0.99776 & 0.00000 & 0.0 & 0.0 \\
5& 0.99956 & 0.00044 & 0.00000 & 0.0 & 0.0 \\
12&0.03497 & 0.96499 & 0.00004 & 0.0 & 0.0 \\
13&0.95128 & 0.04870 & 0.00002 & 0.0 & 0.0
\end{tabular}
\end{table}

\newpage
\begin{figure}
\caption{
The nearest neighbour distributions (NND) of the eigen-spectra for 
$\kappa=1.75$ and 
$\hbar \omega=0.5$ MeV with no Coulomb potential. The NND is calculated for
each isospin bin separately. In all the three cases the NND fits the GOE
distribution.
}
\label{figure.1}
\end{figure}

\begin{figure}
\caption{
The nearest neighbour distribution (NND) of the eigen-spectra for 
$\kappa=1.75$ and 
$\hbar \omega=0.5$ MeV. (a) the results with no Coulomb interaction but
considering all the good isospin states. The calculated NND is between that
expected for GOE and Poissionian distributions. (b) Shows the results
by including the Coulomb potential and the distribution becomes close to
GOE. In (c), the Coulomb matrix elements have been artifically increased
by a factor of 2 and NND fits the GOE distribution with a much 
reduced chi-square.
}
\label{figure.2}
\end{figure}

\begin{figure}
\caption{
The nearest neighbour distributions (NND) of the eigen-spectra for $\kappa=1.75$ and 
$\hbar \omega=0.5$ MeV with Coulomb potential. (a) the NND is calculated
by including the lowest 800 eigen-values, (b) the eigen-values from 801 to
1600 are included and in (c) the eigen-values from 1601 to 2400 are used in
the statistical analysis. The spectra become more chaotic with increasing
excitation energy.
}
\label{figure.3}
\end{figure}

\begin{figure}
\caption{
This is same as Fig. 2 but for $\hbar \omega=2.50$ MeV. The NND appears to deviate
from GOE as compared to Fig. 2. 
}
\label{figure.5}
\end{figure}

\begin{figure}
\caption{
The degree of chaoticity using the measures of Brody (denoted by open circle),
Berry-Robnik (denoted by open square) and spectral-rigidity
(denoted by filled diamond) for $\kappa=1.75$. (a) Shows 
the results with no Coulomb
potential but including all the eigen-values, (b) gives the results with
Coulomb potential and (c) shows the results for T=0 states only.
}
\label{figure.6}
\end{figure}

\begin{figure}
\caption{
The spectral-rigidity parameter of Dyson and Mehta for $\kappa=1.75$.
(a) shows the results with no Coulomb
potential but including all the eigen-values, (b) gives the results with
Coulomb potential and (c) shows the results for T=0 states only.
}
\label{figure.7}
\end{figure}

\begin{figure}
\caption{
This is same as Fig. 5 but with $\kappa=3.50$.
}
\label{figure.8}
\end{figure}

\begin{figure}
\caption{
The overlap-integral, $|<\Phi^{j} (\omega=0.50)|\Phi^{i} (\omega=0.55) >|^2$
in (a), (b) and (c) for i=10, 500 and 1500 using the deformation
$\kappa=1.75$. The overlap of these three
ket states are calculated with all the eigen-states on the bra side, j=1 to
2468. The overlap is maximum for j=i and are shown only around this maximum
value. In (a), the overlap is almost equal to one for j=i and is expected
for a regular spectrum. For the regimes (b) and (c) which are chaotic, 
the overlaps show a broad spectrum.
}
\label{figure.9}
\end{figure}

\begin{figure}
\caption{
The overlap-integral, $|<\Phi^{j} (\omega=2.45)|\Phi^{i} (\omega=2.50) >|^2$
in (a), (b) and (c) for i=10, 500 and 1500 using the deformation
$\kappa=3.50$. The overlap of these three
ket states are calculated with all the eigen-states on the bra side, j=1 to
2468. For all the three cases consided here, the overlap-integral peaks around
j=i and is an indication of the regular motion at all excitation energies
for the present deformation value.
}
\label{figure.10}
\end{figure}


\begin{thebibliography}{9}

\bibitem{gmw98}
T. Guhr, A. Muller-Groeling and H.A. Weidenmuller,
Phys. Rep. {\bf 299} (1998) 189

\bibitem{rei92}
L.E. Reichl,
{\it The Transition to Chaos},
Springer-Verlag, New-York, 1992

\bibitem{bgs84}
O. Bohigas, M.J. Giannoni and C. Schmit,
Phys. Rev. Lett. {\bf 52} (1984) 1

\bibitem{aw91}
Y. Alhassid and N. Whelan,
Phys. Rev. Lett. {\bf 67} (1991) 816

\bibitem{mbes88}
G.E. Mitchell, E.G. Bilpuch, P.M. Endt and J.F. Shriner,
Phys. Rev. Lett. {\bf 61} (1988) 1473

\bibitem{gw90}
T. Guhr and H.A. Weidenmuller,
Ann. Phys. {\bf 199} (1990) 412

\bibitem{ob92}
W.E. Ormand and R.A. Broglia,
Phys. Rev. {\bf C46} (1992) 1710

\bibitem{rud96}
D. Rudolph et al., Phys. Rev. Lett. {\bf 76} (1996) 376

\bibitem{bro81}
T.A. Brody, J. Flores, J.B. French, P.A. Mello, A. Pandey and
S.S.M. Wong, Rev. Mod. Phys. {\bf 53} (1981) 385

\bibitem{be84}
M.V. Berry and M. Robnik, J. Phys. A: Math. Gen. {\bf 17} 
(1984) 2413

\bibitem{meh91}
M.L. Mehta,
{\it Random Matrices},
Academic, New York, 1991

\bibitem{kpr95}
A.T. Kruppa, K.F. Pal and N. Rowley,
Phys. Rev. {\bf C52} (1995) 1826

\bibitem{srn90} 
J.A. Sheikh, M.A. Nagarajan and N. Rowley,
Phys.  Lett. {\bf B240} (1990) 11

\bibitem{she90} 
J.A. Sheikh, N. Rowley, M.A. Nagarajan and H.G. Price,
Phys. Rev. Lett. {\bf 64} (1990) 376

\bibitem{fsr94}  
S. Frauendorf, J.A. Sheikh and N. Rowley,
Phys. Rev. {\bf C50} (1994) 196

\bibitem{swi98}
J.A. Sheikh, D.D. Warner and P. Van Isacker,
Phys. Lett. {\bf B443} (1998) 16

\bibitem{bf79} 
R. Bengtsson and S. Frauendorf, 
Nucl. Phys. {\bf A327} (1979) 139

\bibitem{bo84b}
O. Bohigas and M.-J. Giannoni,  in {\em Mathematical and 
Computational Methods
  in Nuclear Physiscs}, Vol.~209 of {\em Lecture Notes in Physics}, eds.
  J. S. Dehesa, J. M. G. Gomez and A. Polls
  (Springer-Verlag, Berlin, 1984) p. 1

\bibitem{se85a}
T.H. Seligman and J.~J.~M. Verbaarschot, J. Phys. A: Math.
 Gen. {\bf 18},
  2227  (1985)

\bibitem{pr86}
S.T.W.H.~Press, B.P.~Flannery and W. Vettering, 
{\em Numerical Recipes}
  (Cambridge University Press, Cambridge, 1986)

\bibitem{hs} 
K. Hara and Y. Sun, 
Int. J. Mod. Phys. {\bf E4} (1995) 637

\end{thebibliography}
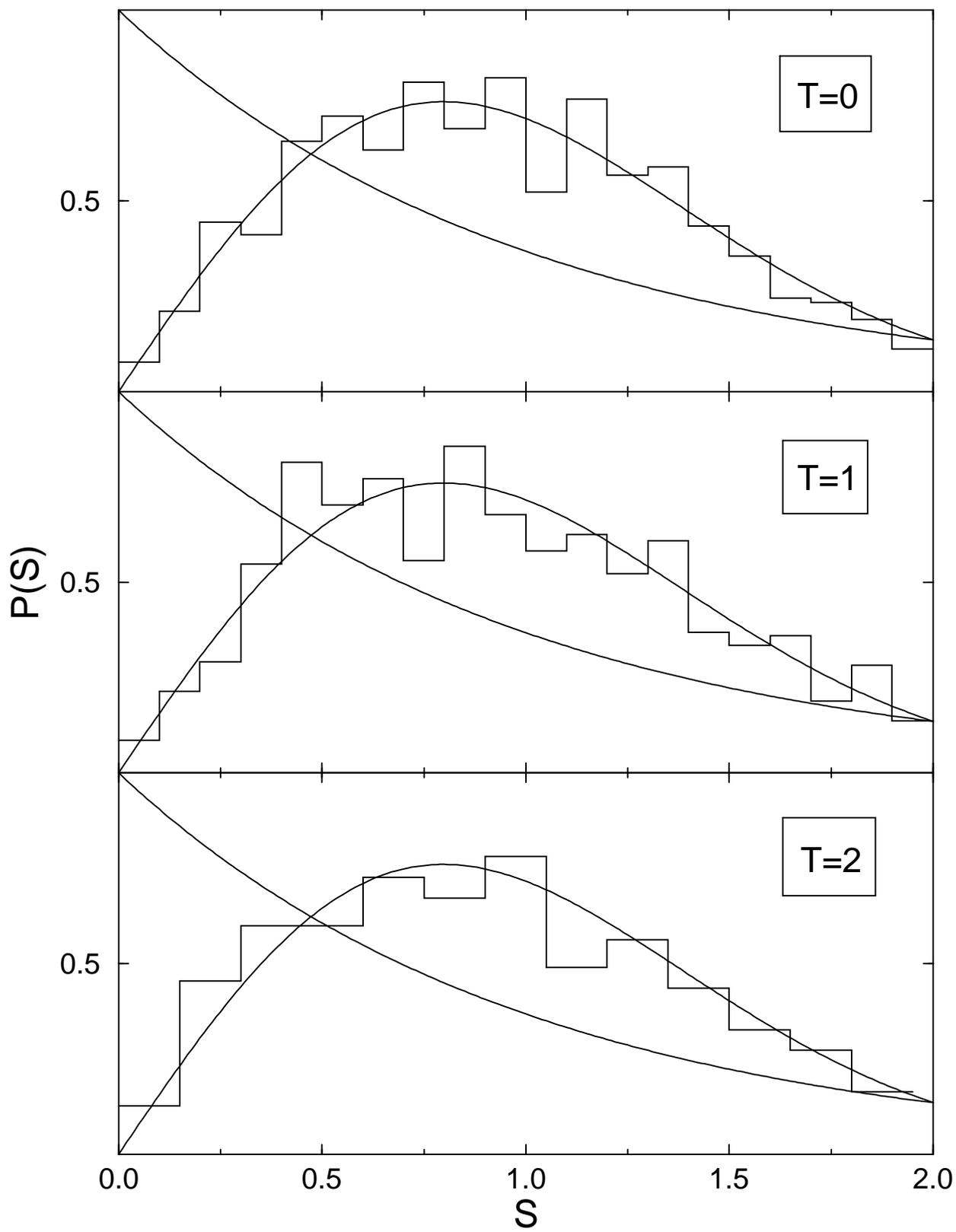
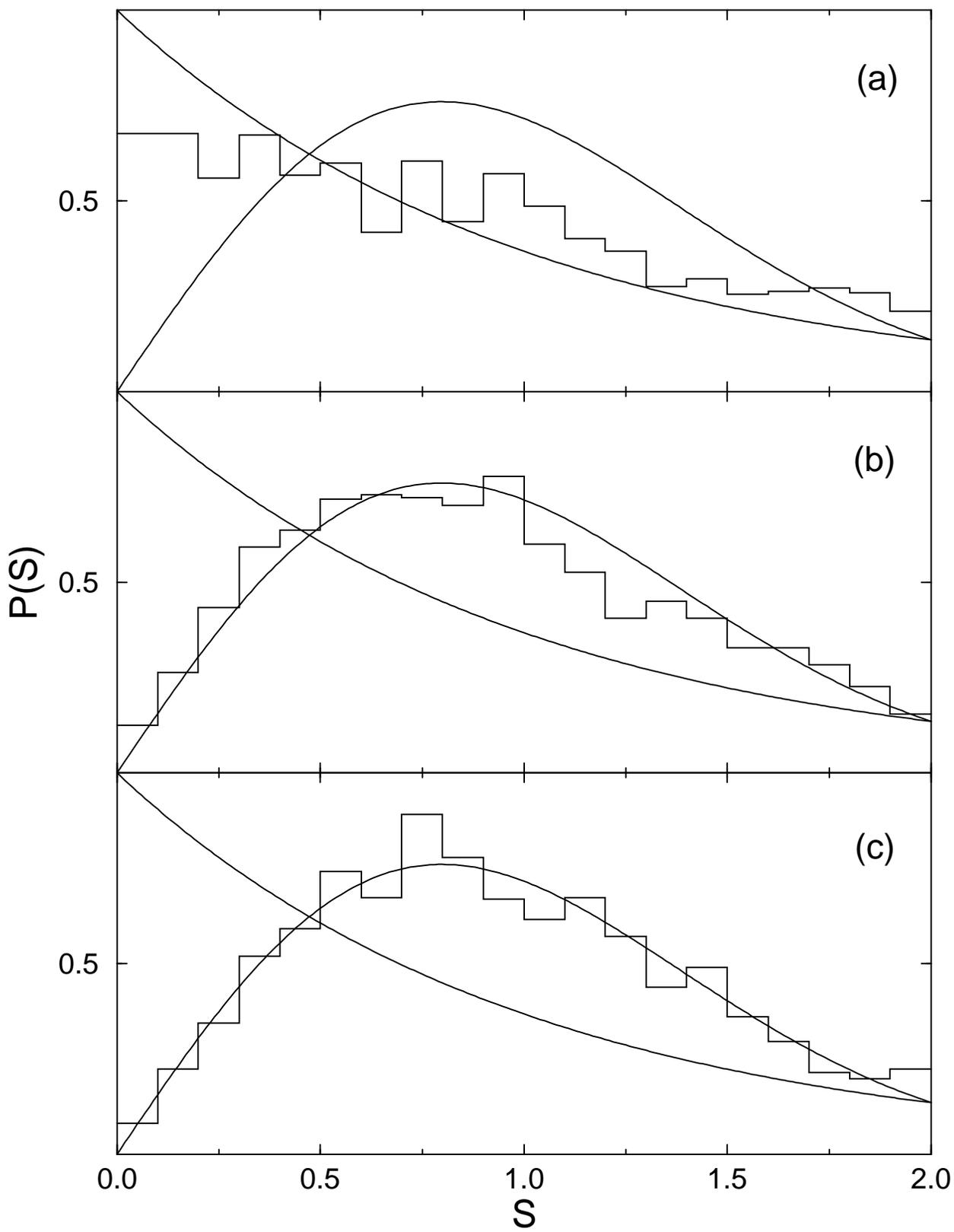
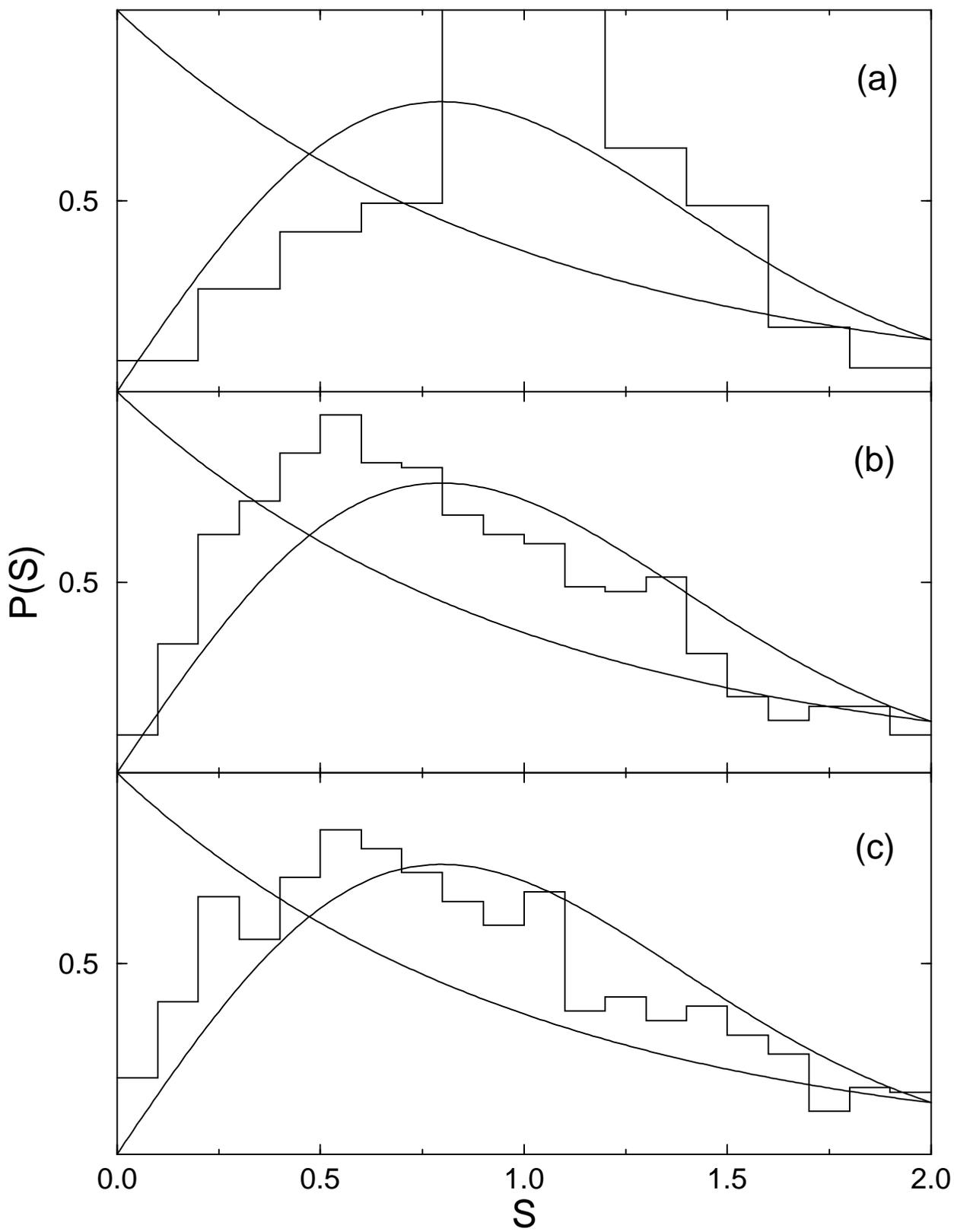
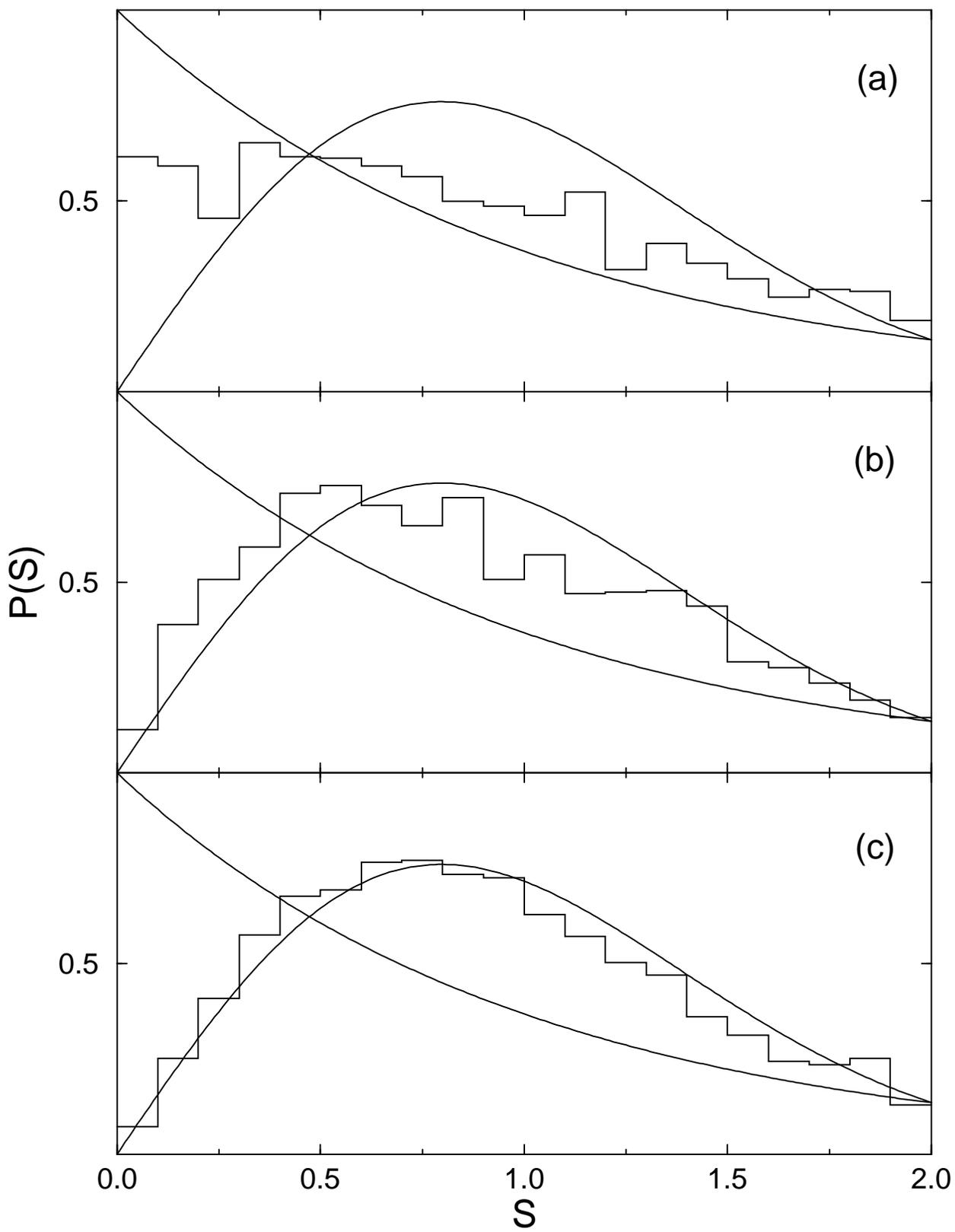
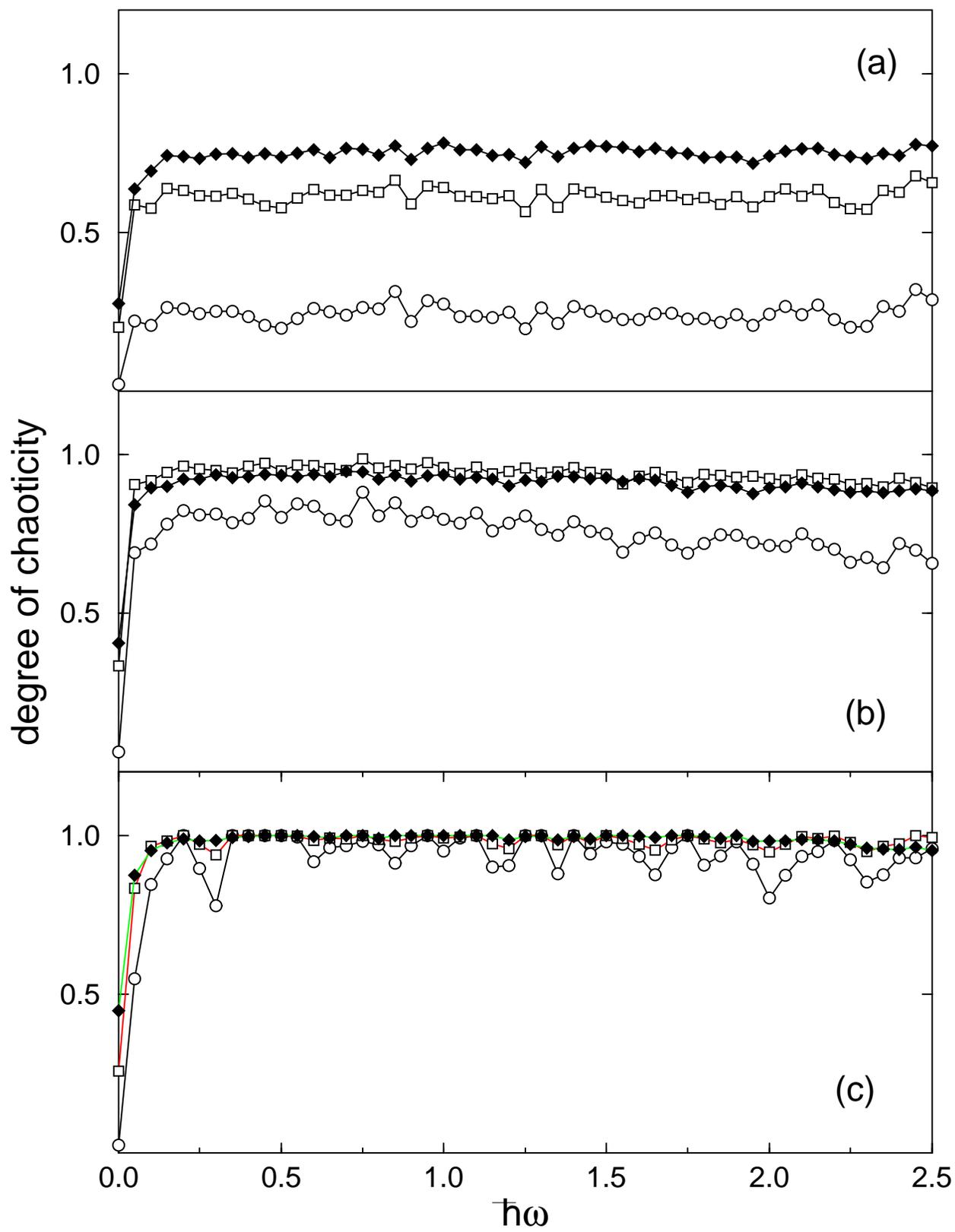
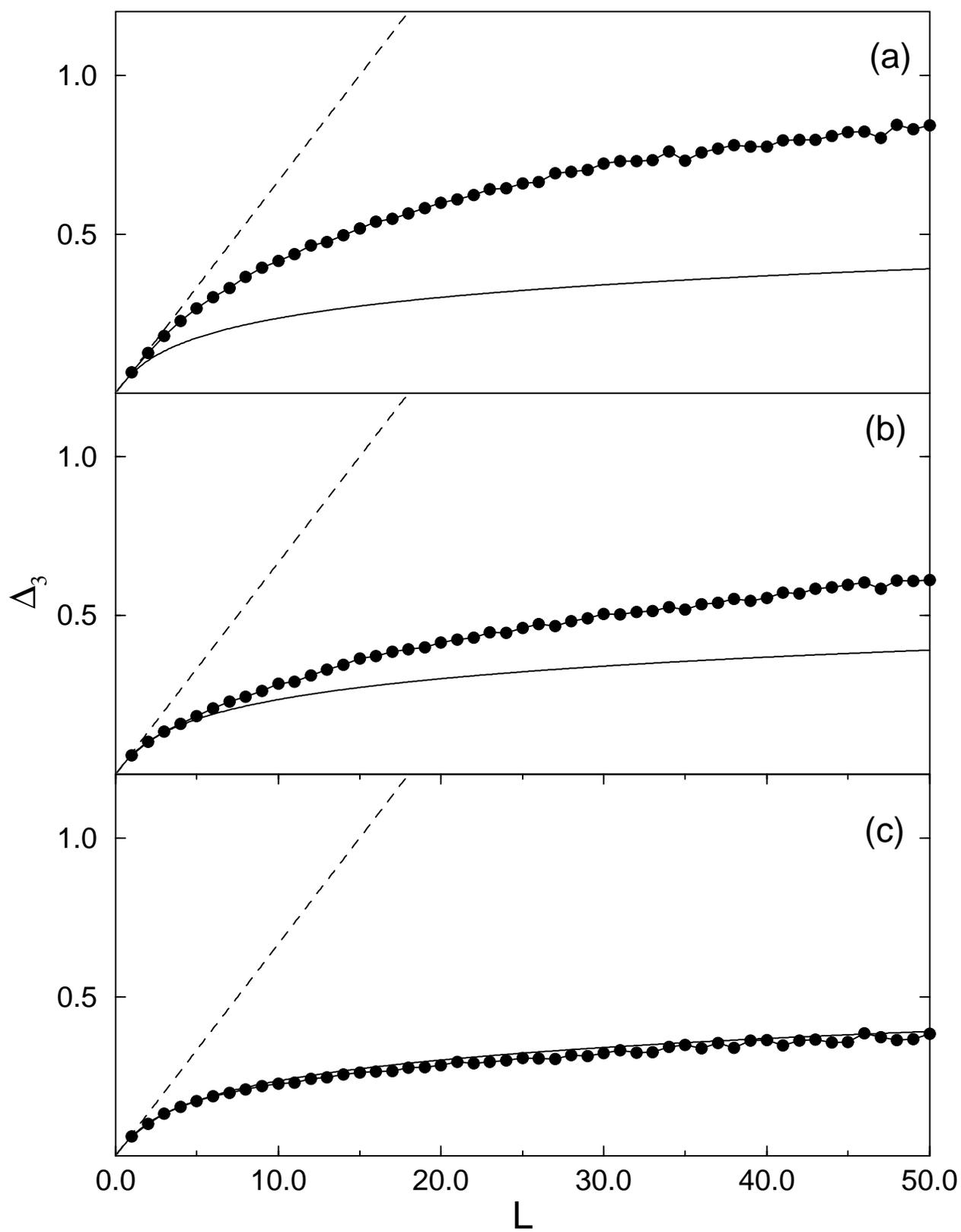
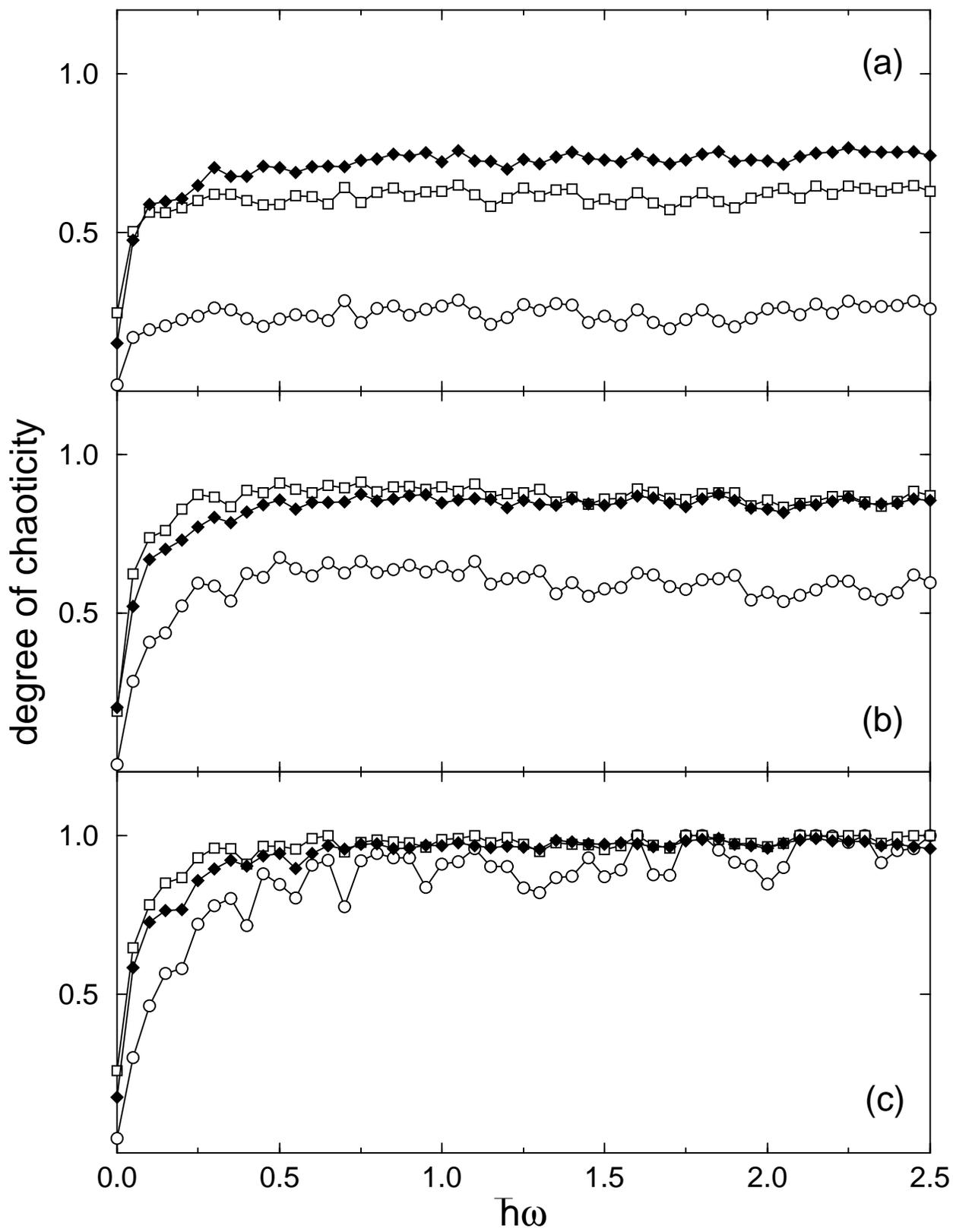
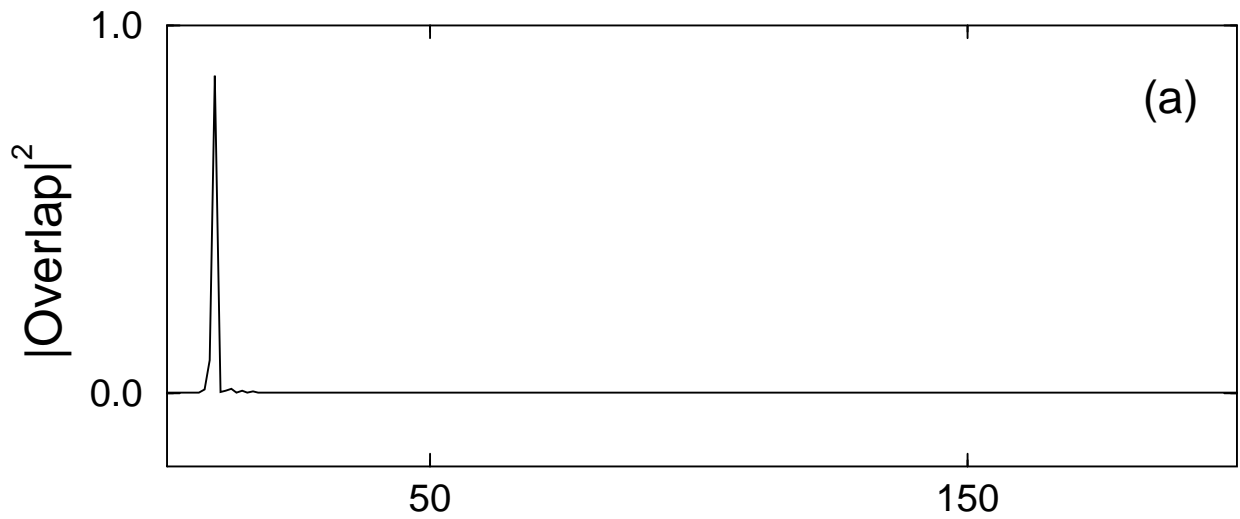
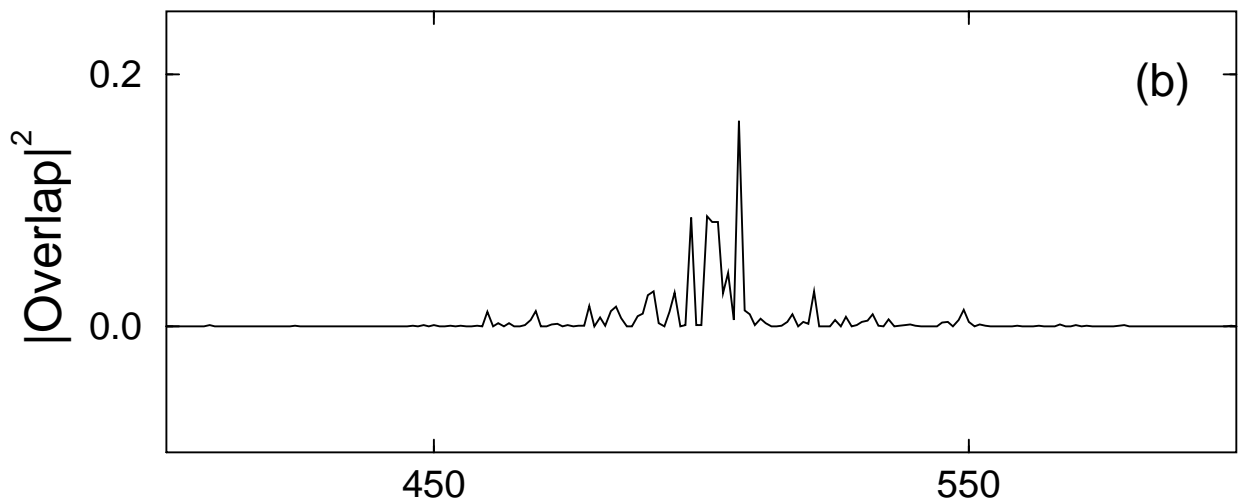
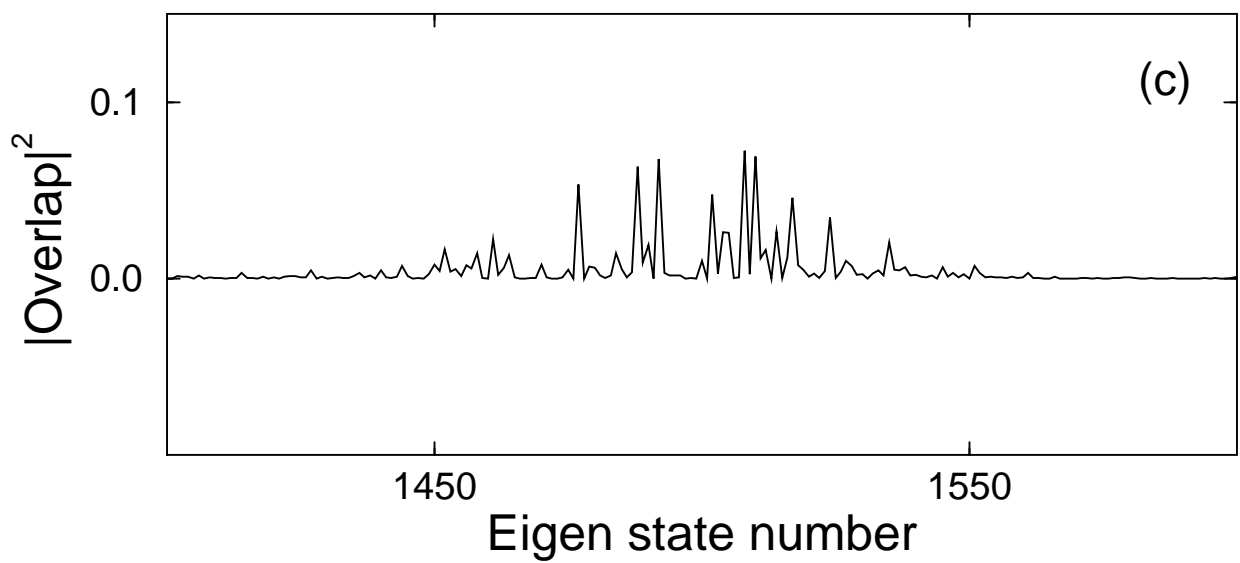
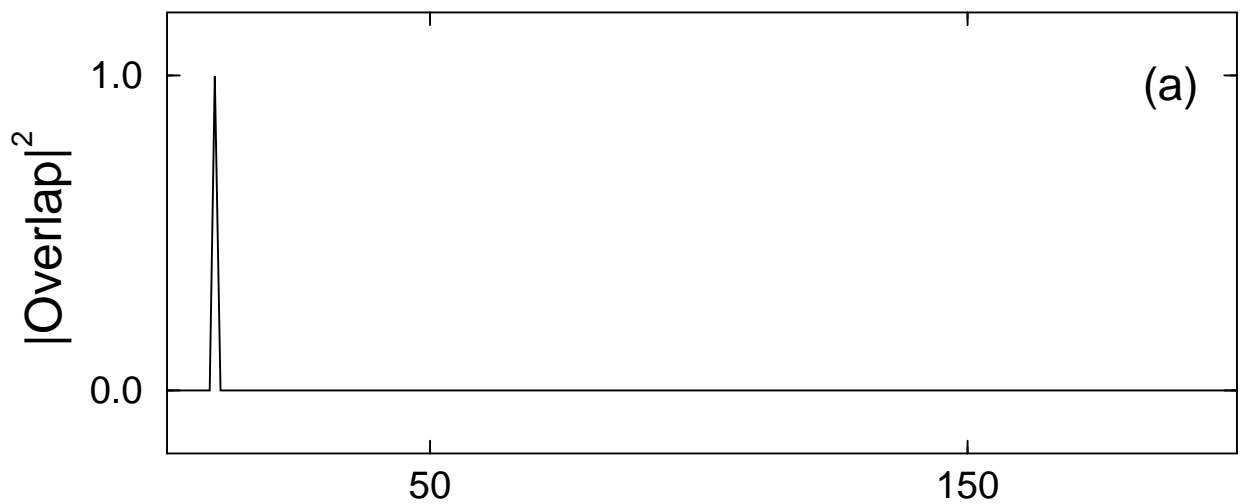
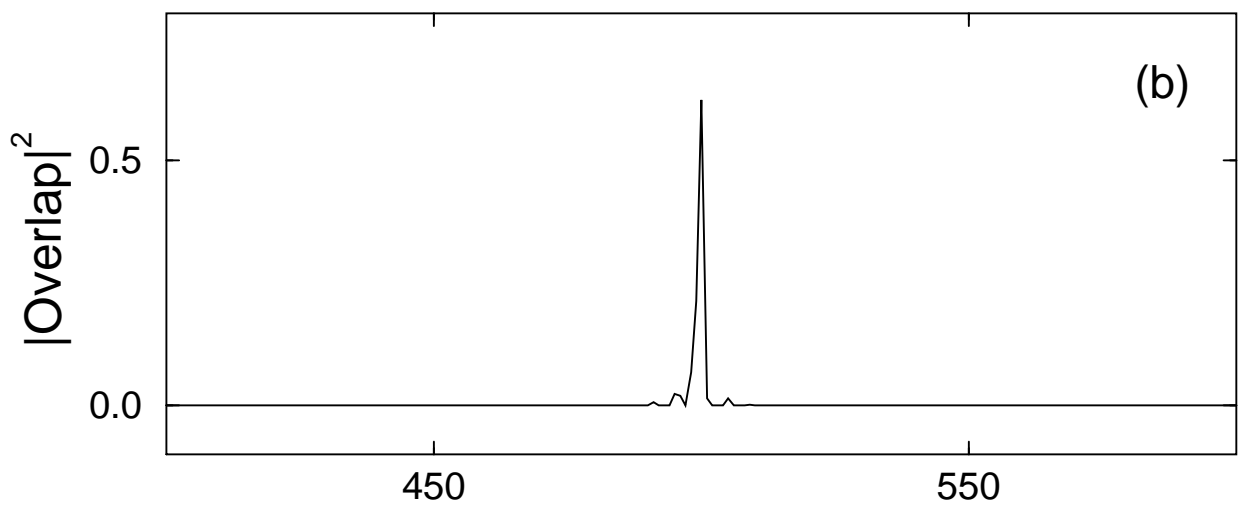
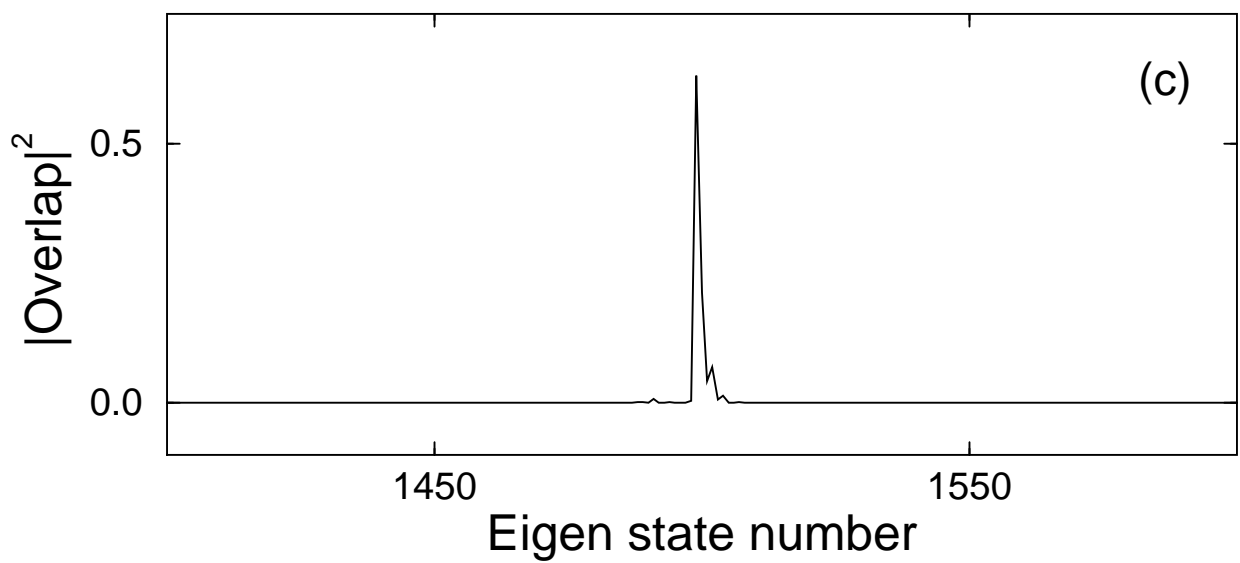
\end{document}